\documentclass[aps,prl,twocolumn,groupedaddress,nofootinbib,showpacs]{revtex4}
\usepackage[dvips]{graphicx}
\newif\ifdraft
\drafttrue
\newif\ifpreprint
\preprinttrue

\def\fig#1{fig.~{\ref{#1}}}

\def\eqn#1{eq.~({\ref{#1}})}
\def\nn{\nonumber}
\def\NeqFour{\mathcal{N}=4}
\def\NeqEight{\mathcal{N}=8}

\def\tab#1{table~{\ref{#1}}}

\def\e{\epsilon}
\def\eps{\epsilon}

\def\tree{{\rm tree}}

\newbox\charbox
\newbox\slabox
\def\s#1{{      
        \setbox\charbox=\hbox{$#1$}
        \setbox\slabox=\hbox{$/$}
        \dimen\charbox=\ht\slabox
        \advance\dimen\charbox by -\dp\slabox
        \advance\dimen\charbox by -\ht\charbox
        \advance\dimen\charbox by \dp\charbox
        \divide\dimen\charbox by 2
        \raise-\dimen\charbox\hbox to \wd\charbox{\hss/\hss}
        \llap{$#1$}
}}

\begin{document}

\title{
\ifpreprint
\hbox{\rm \small CERN--PH--TH/2012/041
$\null$ \hskip 11 cm \hfill  UCLA/12/TEP/101} 
\fi
Absence of Three-Loop Four-Point Divergences in $\mathcal{N} = 4$ Supergravity
}
 
\author{Zvi~Bern${}^{a,b}$, Scott~Davies${}^a$, Tristan~Dennen${}^a$, and 
Yu-tin~Huang${}^{a,c}$}

\affiliation{
${}^a$Department of Physics and Astronomy, UCLA, Los Angeles, CA
90095-1547, USA \\
${}^b$Theory Division, Physics Department, CERN, CH--1211 Geneva 23, 
    Switzerland \\
${}^c$School of Natural Sciences, Institute for Advanced Study, Princeton, NJ 
08540, USA}

\begin{abstract}
We compute the coefficient of the potential three-loop divergence in
pure $\mathcal{N}=4$ supergravity and show that it vanishes, contrary
to expectations from symmetry arguments.  The recently uncovered
duality between color and kinematics is used to greatly streamline the
calculation.  We comment on all-loop cancellations hinting at further
surprises awaiting discovery at higher loops.
\end{abstract}

\pacs{04.65.+e, 11.15.Bt, 11.30.Pb, 11.55.Bq \hspace{1cm}}

\maketitle


Recent years have seen a resurgence of interest in the possibility
that certain supergravity theories may be ultraviolet finite.  This
question had been carefully studied in the late 70's and early 80's in
the hope of using supergravity to construct fundamental theories of
gravity. The conclusion of these early studies was that
nonrenormalizable ultraviolet divergences would almost certainly
appear at a sufficiently large number of quantum loops, though this
remains unproven.  Although supersymmetry tends to tame the
ultraviolet divergences, it does not appear to be sufficient to
overcome the increasingly poor ultraviolet behavior of gravity
theories stemming from the two-derivative coupling.  The consensus
opinion from that era was that all pure supergravity theories would
likely diverge at three loops (see
e.g. ref.~\cite{ThreeLoopPrediction}), though with assumptions,
certain divergences are perhaps delayed a few extra loop
orders~\cite{GrisaruSiegel}.

More recently, direct calculations of divergences in supergravity
theories have been carried out~\cite{BDDPR, GravityThree,
  GravityFour,ck4l}, shedding new light on this issue.  From these
studies we now know that through four loops maximally supersymmetric
$\NeqEight$ supergravity is finite in space-time dimensions, $D < 6/L
+ 4$ for $L=2,3,4$ loops.  These calculations also tell us that the
bound is saturated.  In $D=4$, $E_{7(7)}$ duality
symmetry~\cite{N8Sugra} has recently been used to imply ultraviolet
finiteness below seven loops~\cite{SevenLoopE7}, also explaining the
observed lack of divergences. In a parallel development, string theory
and a first quantized formalism use supersymmetry considerations to
arrive at similar conclusions~\cite{NineLoopRetraction}.  The latter
approach leads to $D$-dimensional results consistent with the explicit
calculations through four loops, but predicts a worse behavior
starting at $L=5$.  At seven loops, the potential four-graviton
counterterm of $\NeqEight$ supergravity~\cite{HoweLindstrom} appears
to be consistent with all known symmetries~\cite{SevenLoopE7,
  VanishingVolume}.  (See ref.~\cite{Kallosh} for a more optimistic
opinion.)  More generally,
$1/{\mathcal{N}}$-BPS operators serve as potential counterterms for
$\mathcal{N}=4,5,6,8$ supergravity at $L=3,4,5,7$ loops, respectively,
suggesting that in $D=4$ ultraviolet divergences will occur at these
loop orders in these theories~\cite{VanishingVolume}.  It therefore
might seem safe to conclude that $\NeqFour$
supergravity~\cite{N4Sugra} in particular will diverge at three loops.

On the other hand, studies of scattering amplitudes suggest that
additional ultraviolet cancellations will be found beyond these.  We
know that even pure Einstein gravity at one loop exhibits remarkable
cancellations as the number of external legs
increases~\cite{UnexpectedOneLoop}.  Through unitarity, such
cancellations feed into nontrivial ultraviolet cancellations at {\it
  all} loop orders~\cite{Finite}.  In addition, the proposed
double-copy structure of gravity loop amplitudes~\cite{BCJLoop}
suggests that gravity amplitudes are more constrained than 
symmetry considerations suggest.  In this Letter we show that the
ultraviolet properties of $\NeqFour$ supergravity are indeed better than 
had been anticipated.

\begin{figure}[t]
\includegraphics[clip,scale=0.35]{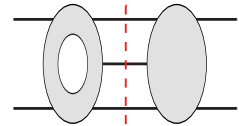}
\caption[a]{A sample cut at three loops 
displaying cancellations in $\NeqFour$ supergravity special to four dimensions.}
\label{CutArgumentFigure}
\end{figure}

To motivate the possibility of hidden cancellations in $\NeqFour$
supergravity, consider the unitarity cut displayed in
\fig{CutArgumentFigure} isolating a one-loop subamplitude in a
three-loop amplitude.  As noted in
refs.~\cite{UnexpectedOneLoop,DunbarEttle}, at one loop a five-point
diagram in an $\NeqFour$ supergravity amplitude effectively can have
up to five powers of loop momenta in the numerator, similar to the
power counting of pure Yang-Mills theory.  There are also three
additional powers of numerator loop momentum coming from the tree
amplitude on the right-hand side of the cut, giving a total of at
least eight powers of numerator loop momentum.  Taking into account
three loop integrals and ten propagators suggests that this amplitude
should diverge at least logarithmically in $D=4$.  (The power counting
analysis of this cut performed in ref.~\cite{DunbarEttle} assumed
that additional powers of numerator loop momenta coming from the tree
amplitude in the cut can be ignored, contrary to our analysis.)

However, this type of power counting is too na\"ive and does not
account for the special property that no one- and two-loop ultraviolet
divergences are present in $D=4$~\cite{Supergravity}. Thus in $D=4$
there are additional cancellations of the loop momenta in one-loop
subdiagrams, effectively removing powers of loop momenta from the
numerators of the loop integrands once all pieces have been combined
and integrated.  These additional cancellations can affect the
higher-loop effective overall power counting.  We show this occurs by
computing the coefficient of the potential three-loop four-point
divergences in $\NeqFour$ supergravity.

\begin{figure}[t]
\includegraphics[clip,scale=0.43]{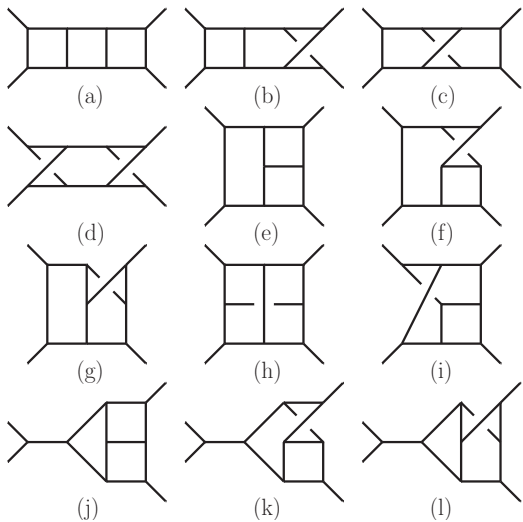}
\caption[a]{The twelve graphs appearing in the three-loop $\NeqFour$
  sYM amplitude~\cite{BCJLoop} and in gravity amplitudes obtained
  from these using the double-copy formula.}
\label{TwelveGraphsFigure}
\end{figure}


To make the calculation of the potential three-loop divergence
feasible, we use the duality between color and kinematics uncovered by
Carrasco, Johansson and one of the authors (BCJ)~\cite{BCJ,BCJLoop}.
According to this conjecture, we can reorganize a (super) Yang-Mills
amplitude into graphs where the numerators satisfy identities in
one-to-one correspondence with color Jacobi identities.  Whenever this
is accomplished, we obtain corresponding gravity amplitudes simply by
replacing color factors with kinematic numerators of a corresponding
second gauge-theory amplitude.  That is, gravity loop amplitudes are given
by~\cite{BCJLoop},
\begin{eqnarray}
 \frac{(-i)^{L+1}}{(\kappa/2)^{n-2+2L}}\mathcal{M}^{\text{loop}}_n = 
    \sum_j \int \prod_{l=1}^L \frac{d^Dp_l}{(2\pi)^D} \frac{1}{S_j}
 \frac{n_j \tilde{n}_j}{\prod_{\alpha_j} p_{\alpha_j}^2}\,, \quad
\label{DoubleCopy}
\end{eqnarray}
where $n_j$ and $\tilde{n}_j$ are kinematic numerator factors from
gauge-theory amplitudes and $\kappa$ is the gravitational coupling.
The factors $S_j$ are the usual combinatoric factors associated with
the symmetries of the graphs.  The sum runs over all distinct
graphs with cubic vertices, such as the ones appearing in
\fig{TwelveGraphsFigure}. The propagators appearing in \eqn{DoubleCopy}
are the ordinary propagators corresponding to the internal lines of
the graphs.  Depending on the particular theory under consideration,
we use different component gauge-theory numerators in
\eqn{DoubleCopy}.

In our study of pure $\NeqFour$ supergravity with no matter
multiplets~\cite{N4Sugra}, we take one component gauge theory to be
$\NeqFour$ super-Yang-Mills (sYM) theory and the second component to
be nonsupersymmetric pure Yang-Mills theory.  This construction was
used in earlier one- and two-loop studies of $\mathcal{N}\ge 4$
supergravity amplitudes~\cite{OneTwoLoopN4}. The main differences in
our case are that integrated gauge-theory expressions are not known
and that the $\mathcal{N} = 4$ sYM numerators are not all independent
of loop momenta.

As explained in refs.~\cite{BCJLoop,BCJSquare}, only one of the two
component gauge-theory amplitudes needs to be in a form manifestly
satisfying the duality for the double-copy property (\ref{DoubleCopy})
to hold.  The other gauge-theory amplitude can be any convenient
representation arranged into diagrams with only cubic vertices.  We
note that our construction applies immediately to all four-point
amplitudes of pure $\NeqFour$ supergravity, since these
are constructed simply by considering all possible external states on
the $\NeqFour$ sYM side of the double copy; at four-points this
information is entirely encoded in an overall prefactor of the tree
amplitude.  At three loops, we take the $\NeqFour$ sYM copy from
ref.~\cite{BCJLoop}, since it has BCJ duality manifest. This
representation of the $\NeqFour$ sYM amplitude is described by
the 12 graphs in \fig{TwelveGraphsFigure}.  For the pure Yang-Mills
copy, we use ordinary Feynman diagrams in Feynman gauge, including
ghost contributions. The contact contributions are assigned to
diagrams with only cubic vertices according to their color factor.  In
this construction, most Feynman diagrams are irrelevant because in the
double-copy formula they get multiplied by vanishing $\NeqFour$
sYM diagram numerators. This construction gives the complete 
three-loop four-point integrand of $\NeqFour$ supergravity.
We have also applied these ideas to reproduce the absence of one- and
two-loop divergences in pure $\NeqFour$ supergravity, starting from
the one- and two-loop four-point sYM amplitudes~\cite{GSB,BRY}.

To prove the correctness of our construction, we use the unitarity
method~\cite{UnitarityMethod,BDDPR}.  The generalized unitarity cuts
decompose the constructed integrand into sums of products of tree
amplitudes, which match against the values obtained using the
double-copy property at tree level~\cite{BCJLoop,BCJSquare}.  Since
all cuts automatically have the proper values in $D$ dimensions, the
amplitude so constructed is correct.

Inserting the numerators of pure Yang-Mills amplitudes generated by
the Feynman rules into the double-copy formula~(\ref{DoubleCopy})
leads to tens of thousands of high-rank tensor integrals, from which
we must extract the ultraviolet divergences. We do so by expanding in
small external momenta.  This gives vacuum diagrams containing both
infrared and ultraviolet divergences.  To deal with ultraviolet
divergences, we use the four-dimensional-helicity regularization
scheme~\cite{FDH}, since it preserves supersymmetry and has been used
successfully in analogous multiloop pure gluon and supersymmetric
amplitudes.  In this scheme, the number of states remain at their
four-dimensional values. Then at the level of the vacuum integrals we
introduce a uniform mass $m$ to separate the infrared divergences from
the ultraviolet ones. Although ultimately there are no one- and
two-loop ultraviolet divergences in $\NeqFour$ supergravity,
individual integrals generally do contain subdivergences due to their
poor power counting.  To deal with this, we make extensive use of the
observations of ref.~\cite{MarcusSagnotti} to subtract
subdivergences integral by integral. Extractions of ultraviolet
divergences in higher-dimensional $\NeqEight$ supergravity were
discussed recently in refs.~\cite{Neq44np,ck4l}.

\begin{figure}[t]
\includegraphics[clip,scale=0.4]{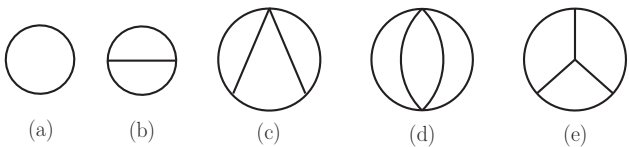}
\caption[a]{The basis of vacuum integrals for one through three loops.}
\label{BasisIntegralsFigure}
\end{figure}

At two or higher loops, the introduced mass regulator 
induces unphysical regulator dependence in
individual integrals. However, at least for logarithmically divergent
integrals, the regulator dependence comes entirely from
subdivergences, which we systematically
subtract~\cite{MarcusSagnotti}. We therefore introduce the mass
regulator before subtracting subdivergences, but after reducing all
integrals to logarithmic by series expansion in small external
momenta.  To implement the subtractions, we recursively define the
subtracted divergence $\mathcal{S}[\ldots]$ of an integral,
\begin{eqnarray}
\label{SubtractedDefinition}
&&  \mathcal{S}\Biggl[\int \prod_{i=1}^{L} dp_i\, I  \Biggr] \equiv
    \text{Div}\Biggl[\int \prod_{i=1}^{L} dp_i\, I \Biggr]  \\
  &&\quad  - \sum_{l=1}^{L-1} \sum_{{l-\text{loop}}\atop{\text{subloops}}} \text{Div}\Biggl[\int \prod_{j=l+1}^{L} dp_j'\,
\mathcal{S}\Biggl[\int \prod_{i=1}^{l} dp_i' \, I\Biggr] \Biggr]\,,
\nn \hskip 1 cm 
\end{eqnarray}
where $\text{Div}[\ldots]$ is the complete divergence of an integral,
$I$ is the integrand, and $dp_i$ is shorthand for
$d^Dp_i/(2\pi)^D$. The sum over subloops must include \textit{all}
subloops of the diagram where a subdivergence could occur --- not just
the loops that are manifestly parametrized by $p_i$ --- and here we
have indicated this by changing variables to $p_i'$ in each
subtraction, such that the $l-$loop subintegral under consideration is
parametrized by $p_1',\ldots,p_l'$. For example, graph (e) in
\fig{TwelveGraphsFigure} has seven one-loop subintegrals and six
two-loop subintegrals to consider, and each two-loop subintegral has
three one-loop subintegrals of its own.

\begin{table}[t]
  \begin{center}
\begin{tabular}[t]{|c|c|}
\hline
Graph & $(\text{divergence})/(\langle 12 \rangle^2 [34]^2 stA^\tree (\frac{\kappa}{2})^8)$ \\
\hline
(a)-(d) & 0 \\
(e) & $\frac{263}{768}\frac{1}{\e^3}+\frac{205}{27648}\frac{1}{\e^2}+\left(-\frac{5551}{768}\zeta_3+\frac{326317}{110592}\right)\frac{1}{\e}$\\
(f) & $-\frac{175}{2304}\frac{1}{\e^3}-\frac{1}{4}\frac{1}{\e^2}+\left(\frac{593}{288}\zeta_3-\frac{217571}{165888}\right)\frac{1}{\e}$\\
(g) & $-\frac{11}{36}\frac{1}{\e^3}+\frac{2057}{6912}\frac{1}{\e^2}+\left(\frac{10769}{2304}\zeta_3-\frac{226201}{165888}\right)\frac{1}{\e}$\\
(h) & $-\frac{3}{32}\frac{1}{\e^3}-\frac{41}{1536}\frac{1}{\e^2}+\left(\frac{3227}{2304}\zeta_3-\frac{3329}{18432}\right)\frac{1}{\e}$\\
(i) & $\frac{17}{128}\frac{1}{\e^3}-\frac{29}{1024}\frac{1}{\e^2}+\left(-\frac{2087}{2304}\zeta_3-\frac{10495}{110592}\right)\frac{1}{\e}$\\
(j) & $-\frac{15}{32}\frac{1}{\e^3}+\frac{9}{64}\frac{1}{\e^2}+\left(\frac{101}{12}\zeta_3-\frac{3227}{1152}\right)\frac{1}{\e}$\\
(k) & $\frac{5}{64}\frac{1}{\e^3}+\frac{89}{1152}\frac{1}{\e^2}+\left(-\frac{377}{144}\zeta_3+\frac{287}{432}\right)\frac{1}{\e}$\\
(l) & $\frac{25}{64}\frac{1}{\e^3}-\frac{251}{1152}\frac{1}{\e^2}+\left(-\frac{835}{144}\zeta_3+\frac{7385}{3456}\right)\frac{1}{\e}$\\
\hline
  \end{tabular}
  \end{center}
\caption{The graph-by-graph divergences for the four-graviton
  amplitude with helicities $(1^-2^-3^+4^+)$ (up to an overall
  normalization). Each expression includes a permutation sum over
  external legs, with the symmetry factor appropriate to the
  graph. These quantities are not individually gauge-invariant, and here we use 
  spinor helicity with the choice of reference momenta $q_1=q_2=k_3$ and
  $q_3=q_4=k_1$.  The sum over the diagram contributions vanishes.}
\label{DiagramResultsTable}
\end{table}

By the time we apply \eqn{SubtractedDefinition}, each integral has a
single scale given by the mass regulator $m$. We are left with the
task of calculating the divergences $\text{Div}[\ldots]$ of
single-scale vacuum integrals. To evaluate these integrals, we first
eliminate tensors composed of loop momenta from the numerators by
noticing that the integrals must be linear combinations of products of
metric tensors $\eta^{\mu\nu}$. (See ref.~\cite{ck4l} for a recent
discussion of evaluating tensor vacuum integrals.)  Then we reduce the
resulting scalar integrals to a basis using integration by parts, as
implemented in {\sc FIRE}~\cite{Fire}.  The resulting basis is given
by the scalar vacuum integrals shown in \fig{BasisIntegralsFigure}
(along with products of lower-loop integrals), with a single massive
propagator corresponding to each line.  As cross checks we also used
{\sc MB}~\cite{MB} and {\sc FIESTA}~\cite{Fiesta}.  We evaluated all
but the last of these integrals analytically to the required order in
$\e$ by Mellin-Barnes integration with resummation of residues using
the methods presented in ref.~\cite{Davydychev}  (see also
refs.~\cite{OtherIntegralPapers}). The last integral
can be evaluated analytically by making a two-loop subintegral
massless and integrating it first. This does not affect the
ultraviolet divergence because there are no subdivergences in this
case. (The value of the two-loop subintegral can be found in
ref.~\cite{Grozin}.) The results are rational linear combinations of
the transcendental numbers $\zeta_2$, $\zeta_3$, and $\sqrt{3}\, {\rm
  Im}\left(\text{Li}_2\left(e^{i\pi/3}\right)\right)$.

At two loops only the first two integrals shown in
\fig{BasisIntegralsFigure} are needed.  Adding together the
contributions reproduces the fact that there are no two-loop
divergences in pure supergravity theories. At three loops all vacuum
integrals in \fig{BasisIntegralsFigure} contribute.  In
\tab{DiagramResultsTable}, we have collected the derived divergences
of the three-loop four-graviton amplitude for each graph in
\fig{TwelveGraphsFigure}.  The results shown in the table are summed
over the independent permutations including symmetry factors.  The
individual graphs are not gauge invariant and are valid only for the
indicated choice of spinor-helicity reference momenta (see e.g.
ref.~\cite{TreeReview}).  We have divided out a prefactor depending on
the four-point color-ordered super-Yang-Mills tree amplitude, spinor
inner products and the usual Mandelstam invariants $s$ and $t$.  We
note that the transcendental numbers except $\zeta_3$ cancel within
each graph. 

In the sum over all contributions (obtained by adding the rows in the
\tab{DiagramResultsTable}), not only do the $1/\eps^3$ and $1/\eps^2$
divergences cancel, as required because there are no divergent
subamplitudes, but the $1/\eps$ singularity also cancels.  This proves
that the three-loop amplitude is ultraviolet finite.  As a rather
nontrivial check, we confirmed that the sum over all contributions is
independent of reference momentum choices.  As another nontrivial
confirmation, we found that by introducing a uniform mass in the
amplitude at the start of the calculation, all ultraviolet divergences
cancel without the need for subdivergence subtraction.  This matches
expectations that all ultraviolet subdivergences should cancel out
from the total amplitude (although there may be potential regulator
dependence issues with this approach).  We have similarly computed
all four-point amplitudes in the theory using formal polarizations
instead of helicity states, demonstrating that there are no
divergences in any of the three-loop four-point amplitudes of the theory.

In summary, we used the recently uncovered duality between color and
kinematics to streamline the calculation of the coefficient of the
potential three-loop ultraviolet divergence of $\NeqFour$
supergravity, proving that it vanishes.  Might cancellations persist
beyond this?  It is interesting to note that the $D=4$ cancellations
found in one- and two-loop subamplitudes and used to motivate our
three-loop computation can be used just as well to argue for
higher-loop cancellations.  Moreover, the double-copy property of
gravity amplitudes shows there is more structure than captured by the
known symmetries.  Our three-loop calculation provides a concrete
example showing that power counting based on known symmetries can be
misleading.  The results of this paper strongly motivate further
high-loop explorations of the ultraviolet divergence structure of
supergravity theories.  In particular, they emphasize the importance
of explicitly computing the ultraviolet properties of $\NeqEight$
supergravity at five loops.


\vskip .2 cm 

We thank C.~Boucher-Veronneau, J.~J.~Carrasco, L.~Dixon, S.~Ferrara,
H.~Ita, H.~Johansson, R.~Kallosh, D.~A.~Kosower, R.~Roiban, W.~Siegel
and K.~Stelle for helpful discussions.  This research was supported by
the US Department of Energy under contract DE--FG03--91ER40662.

\end{document}